\documentclass{ws-procs9x6}

\begin{document}

{\hfill IPPP/09/18, DCPT/09/36}

\title{THE PHENOMENOLOGY OF \\ GRAVITINO DARK MATTER SCENARIOS \\ IN SUPERGRAVITY MODELS}

\author{Y. SANTOSO}

\address{Institute for Particle Physics Phenomenology, \\
Department of Physics, University of Durham, \\
Durham, DH1 3LE, United Kingdom \\
E-mail: yudi.santoso@durham.ac.uk 
}

\begin{abstract}
We review the phenomenology of gravitino dark matter within supergravity framework. Gravitino can be dark matter if it is the lightest supersymmetric particle, which is stable if $R$-parity is conserved. 
There are several distinct scenarios depending on what the next to lightest supersymmetric particle (NLSP) is. We discuss the constraints and summarize the phenomenology of neutralino, stau, stop and sneutrino NLSPs. 
\end{abstract}


\bodymatter

\section{Introduction}\label{sec1:intro}

The identity of dark matter has not yet been resolved at the time this article was written. One interesting possibility is to have a neutral lightest supersymmetric particle (LSP), which is stable via $R$-parity conservation, as the dark matter\cite{EHNOS}. Neutralino is a popular candidate for dark matter and sneutrino is another possibility~\footnote{Left-sneutrino is excluded by direct detection\cite{snuL}, but right-sneutrino is still viable.}. Here, we focus our attention on yet another hypothesis within supergravity, i.e. gravitino as cold dark matter\cite{GDM-CMSSM,otherGDM}.  

In supergravity models (i.e. models with gravity mediated supersymmetry breaking)\cite{SUGRA}, the gravitino mass is close to the other sparticle masses. However, it is not a priori whether the gravitino is lighter or heavier than the others. Note that this is different from gauge mediated models in which the gravitino mass can be naturally very light\cite{GMSB}. We assume supergravity models here, with supersymmetric masses of $\sim 1\,{\rm GeV} - 1$~TeV and the gravitino is the LSP.   
In this framework, the coupling between gravitino and matter fields is very small, $\sim 1/M_{\rm Pl}$. Because of this, gravitino is practically undetectable (aside from its gravitational effect). Also, the next lightest supersymmetric particle (NLSP) could be long lived, with a lifetime of typically O(1\,s) or longer. In this case, the NLSP decay  affects the primordial light element abundances\cite{moroi}.  
The phenomenology of this scenario depends largely on what the NLSP is. 
We will discuss below various possibilities, each with its own distinct phenomenology.

\section{Gravitino Dark Matter in Supergravity Models}

The biggest theoretical uncertainty in supersymmetric models arises from the fact that we do not know how supersymmetry, if it does exist, is broken in nature. Because of this, the values of the soft couplings are uncertain. We can take them as free parameters. However, because of the large number of parameters, we need to make some simplifying assumptions. 
Motivated by the Grand Unified Theory (GUT), the usual assumption is that parameters of the same type are unified at the GUT scale. Their values at weak scale are then derived by employing the renormalization group equation (RGE). 
The simplest model of this kind is the 
CMSSM (Constrained Minimal Supersymmetric Standard Model)~\footnote{Also known as mSUGRA, depending on your preference.}, in which we have 
universal gaugino mass $m_{1/2}$, universal sfermion mass $m_0$, and
universal trilinear coupling $A_0$ at the GUT scale. In addition, we have two parameters from the Higgs sector, i.e. the ratio of the two Higgs vevs $\tan \beta\equiv \langle H_1\rangle/\langle H_2 \rangle$, and the sign of $\mu$ where $\mu$ is the Higgs mixing parameter in the superpotential.
Note however that in GUT theories, e.g. $SU(5)$ or $SO(10)$, the Higgs fields are contained in different multiplets as compared to the matter multiplets. This motivates a generalization of the CMSSM, in which the Higgs soft masses $m_{1,2}$ are not necessarily equal to $m_0$ at the GUT scale\cite{NUHM}. Furthermore, we can trade  $m_{1,2}$ with $\mu$ and the CP-odd Higgs mass $m_A$ as our free parameters through the electroweak symmetry breaking condition. The resulting model is called Non-Universal Higgs Masses (NUHM) model\cite{ourNUHM}. Thus the NUHM parameters are $m_{1/2}$, $m_0$, $A_0$, $\tan \beta$, $\mu$ and $m_A$.

In the usual scenario with neutralino LSP, we implicitly assume that gravitino is sufficiently heavy such that it decouples from the low energy theory. In the gravitino dark matter (GDM) scenario, on contrary, we assume that the gravitino mass $m_{\widetilde{G}} = m_{3/2}$ is sufficiently small such that the gravitino is the LSP. For our purposes, we can take $m_{3/2}$ as another free parameter. Within CMSSM, with gravitino LSP, there are three possible NLSP, i.e. neutralino, stau and stop particles\cite{GDM-CMSSM,stopNLSP}. For NUHM, in addition, we can have selectron or sneutrino as the NLSP\cite{GDM-NUHM}. Of course for a more general model of MSSM we can have more possibilities.

\section{Phenomenological Constraints}

\subsection{Dark matter relic density constraint}

Being a very weakly interacting particle, gravitino decoupled very quickly from the thermal plasma in the early universe. This leads to a concern that the gravitino could be over-abundance. However, inflation can solve this problem\cite{ELN}. In inflationary models, the early gravitino density together with other densities are diluted by the inflation. Gravitino is then reproduced by reheating after the inflation\cite{KL}, although with a smaller yield that can still satisfy the relic density constraint~\footnote{This can still impose a strong constraint on the inflationary theories~\cite{graInf}. However, this topic is beyond the scope of this article.}.  

Gravitino relic density consists of two parts, the thermal relic $\Omega_{\widetilde{G}}^{\rm T}$ which is produced by reheating, and the non-thermal relic $\Omega_{\widetilde{G}}^{\rm NT}$ coming from the decay of the NLSP. 
\begin{equation}
\Omega_{\widetilde{G}} h^2 = \Omega_{\widetilde{G}}^{\rm T} h^2  + \Omega_{\widetilde{G}}^{\rm NT} h^2
\label{eq:relic}
\end{equation}
The thermal relic is related to the reheating temperature $T_R$ through the following relation~\cite{BBB}  
\begin{equation}
\Omega_{\widetilde{G}}^{\rm T} h^2  \simeq 0.27 \left( \frac{T_R}{10^{10}\, {\rm GeV}} \right) \left( \frac{100 \, {\rm GeV}}{m_{\widetilde{G}}} \right) \left( \frac{m_{\tilde{g}}}{1 \, {\rm TeV}} \right)^2 
\end{equation}
where $m_{\tilde{g}}$ is the gluino mass. We can see that for $m_{\widetilde{G}} = 100$~GeV and $m_{\tilde{g}} = 1$~TeV, to get $\Omega_{\widetilde{G}}^{\rm T} h^2 \lesssim 0.1$ we need $T_R \lesssim 10^{10}$~GeV. The value of $T_R$ depends on the inflation model. For our purpose, we take $T_R$ as a free parameter. 
For one to one decays of NLSP to gravitino, which is generally the case, the gravitino non-thermal relic can be written as:
\begin{equation}
\Omega_{\widetilde{G}}^{\rm NT} h^2 = \frac{m_{\widetilde{G}}}{m_{\rm NLSP}} \Omega_{\rm NLSP} h^2
\end{equation}
where $\Omega_{\rm NLSP} h^2$ is the NLSP density before the decay.
Due to the long lifetime, the NLSP density is frozen out long before its decay, and this can be calculated by the
usual method of solving the Boltzmann equation in the expanding universe. Note that even if the NLSP density is larger than the WMAP value, we might still satisfy the dark matter relic density constraint because of the rescaling by the mass ratio $m_{\widetilde{G}}/m_{\rm NLSP}$.   
The NLSP density can also be written in term of the yield, $Y_{\rm NLSP} = n_{\rm NLSP}/s$, where $n$ is the number density and $s$ is the entropy density. 
This is related to $\Omega_{\rm NLSP} h^2$ by 
\begin{equation}
Y_{\rm NLSP} M_{\rm NLSP} = \Omega_{\rm NLSP} h^2 \times (3.65 \times 10^{-9} \; {\rm GeV}) 
\end{equation}

The total relic density of the gravitino must not exceed the upper limit of the dark matter relic density range as suggested by WMAP\cite{WMAP5}:
\begin{equation}
\Omega_{\rm DM}^{\rm WMAP} h^2 \simeq 0.113 \pm 0.004
\end{equation}
Taking $2 \sigma$, this means that $\Omega_{\widetilde{G}} h^2 \lesssim 0.121$. The percentage of the thermal versus non-thermal relic density depends on the strength of the NLSP interactions which determine $\Omega_{\rm NLSP} h^2$.  If we can measure these interactions at colliders we can deduce the reheating temperature\cite{TR}.

\subsection{The BBN constraints}

Big bang nucleosynthesis (BBN) is often cited as the greatest success of the big bang theory. By using simple assumptions that the early universe is in thermal equilibrium and expanding one can calculate the primordial light element abundances (using standard nuclear cross sections) and gets results which agree very well with the observations. 
If there is a metastable particle that decays during or after the BBN era, the
light element abundances can be altered by the participation of the energetic decay products in the nucleosynthesis processes. Thus, BBN provides a stringent constraint for gravitino dark matter. On the other hand, the prediction of the standard BBN (sBBN) is not perfectly in agreement with the observational data. There seems to be discrepancy between the observed lithium abundances and the predicted values as shown in Table~\ref{li-tbl}.  
\begin{table}
\tbl{Comparison of lithium abundances from standard model prediction and observations.}
{\begin{tabular}{lcc}
\toprule
& $^7$Li/H & $^6$Li/$^7$Li \\
\colrule
Observation & $(1-2)\times 10^{-10}$ & $\sim 0.01 - 0.15$ \\
Standard BBN (with CMB) & \hphantom{00}$\sim 4 \times 10^{-10}$ & $< 10^{-4}$ \\
\botrule
\end{tabular}}
\label{li-tbl}
\end{table} 
This is known as the lithium problem. The discrepancy on $^6$Li is particularly difficult to be solved within the standard theory. There is no known astrophysical process that can produce large amount of $^6$Li. Moreover $^6$Li is fragile. Therefore we should expect less rather than more $^6$Li compared to the prediction. 
The lithium problem could be an indication of a new physics beyond the standard model. There are two proposed solution to this problem through a hypothesized metastable particle. The first one is through catalytic effect\cite{pospelov,otherCat}. 
The process $d + ^4$He  $\to \, ^6$Li $+ \gamma$ is suppressed by parity.  
If there is a massive negatively charged particle $X^-$ that is bound to $^4$He by Coulomb interaction it can absorb the emitted photon, hence the process is no longer parity suppressed. Simultaneously, the $X^-$ particle is freed from the bound by the energy released and can subsequently be attached to another $^4$He, thus acting as a catalysis for $^6$Li production. 
This catalytic effect can also effect other light element abundances such as beryllium\cite{beryCat}. 

Another proposed solution to the lithium problem is through hadronic decay of a metastable particle\cite{jedamzik}. The decay produces energetic $n$, $p$ and also T, $^3$He (through spallation of $^4$He), which then interact with the ambient nuclei, e.g. $n + p \to $D, T + $^3$He $\to ^6$Li (producing more $^6$Li), and $^7$Be($n$,$p$)$^7$Li($p$,$^4$He)$^4$He (reducing $^7$Li). Note that deuterium is also produced, hence put some constraints on this scenario.

\subsection{Astrophysical constraints}

If the NLSP decays at a time later than the BBN, the photons produced by the decay might not be fully thermalised by the time of recombination. This can cause distortion on the cosmic microwave background radiation black body spectrum\cite{lamon+durrer}. 
The size of the distortion depends on the amount of the energy injected into photons. This is represented by a chemical potential $\mu$ for photon.
CMB spectrum measurement by the FIRAS instrument onboard of COBE satellite sets an upper limit on $\mu$\cite{COBE-FIRAS}:
\begin{equation}
\left| \mu \right| \lesssim 9 \times 10^{-5} 
\end{equation}
This limit is only important for lifetime $\gtrsim 10^6$~s since photons produced earlier should have enough time to thermalize before recombination. 

Gravitinos from the NLSP decay at a late time have larger velocities compared to the primordial gravitino. This leads to a longer free-streaming length, smoothing out small scale density perturbation. If the dark matter relic density is dominated by non-thermal relic, the structure formation would be affected. This scenario is proposed as a solution to the small scale problem\cite{warmgravitino}.

\subsection{Collider constraints}

At colliders, heavier supersymmetric particles can still be produced provided that there is enough energy in the collisions. These sparticle would quickly decay, cascading down to the NLSP. Due to its long lifetime the NLSP itself would escape from the detector before eventually decays to gravitino, hence appear as a stable particle with respect to the detectors. There have been searches for stable massive particles (SMP) at colliders\cite{fairbairn}. 

A particularly interesting signal would be produced if the NLSP is electromagnetically charged.  In this case the NLSP should traverse the calorimeter  and subsequently be detected by the muon detector. The first obvious step of the data analysis is of course to discover this charged NLSP.
 The CDF collaboration, based on 1.0~fb$^{-1}$ of data at $\sqrt{s} = 1.96$~TeV, sets a lower bound on (meta)stable stop particle at $249$~GeV\cite{CDF-CHAMPS}; while the D0 collaboration, using 1.1~fb$^{-1}$ of data, sets upper limits for stable stau pair production cross section from 0.31~pb to 0.04~pb for stau masses between 60~GeV and 300~GeV\cite{D0-CHAMPS} and lower mass limits of 206~GeV and 171~GeV for pair produced stable charged gauginos and higgsinos respectively. 

However, not all possible NLSP are charged. If neutralino or sneutrino is the NLSP, they would not be detected. Only neutralino with a very short lifetime ($\lesssim$ few ns) can be detected through its decay product. The CDF sets a lower limit on the neutralino mass at 101~GeV for lifetime 5~ns\cite{CDF-chi}. 
Similar to the familiar case with neutralino LSP, there are also various signatures from the cascade decays. The same methods of analysis can be applied to the case with stable neutral NLSP. For long-lived neutralino NLSP the signatures would be indistinguishable from that of neutralino LSP scenario. For sneutrino NLSP, however, the signatures would in general be different\cite{CoKra}.

\section{Phenomenology of GDM with Various NLSP} 

In this section we look at each scenario of neutralino, stau, stop and sneutrino NLSP. We do not include chargino NLSP\cite{charginoNLSP} here.

\subsection{Neutralino NLSP}

For neutralino mass of 1~TeV and gravitino mass of 1~GeV the neutralino lifetime is about $O(1)$~s. The lifetime is longer for a smaller mass gap. Thus the neutralino in this scenario would escape the collider detectors and trigger large missing energy signatures. Assuming that all primordial neutralinos has eventually decayed to gravitino, only gravitino is floating around today. Therefore WIMP direct detection experiments would not see any signal. There would be no indirect astrophysical signal from dark matter annihilation in the halo either. 
In this case it would be difficult to proof the identity of the dark matter, i.e. whether it is gravitino, axino or whether something else. We will also need to check whether $R$-parity is really conserved. 

The neutralino NLSP in the CMSSM is much constrained by the BBN, especially when  the neutralino lifetime is $\gtrsim 10^4$~s where the electromagnetic shower effect on the light element abundances becomes important. The main reason is because neutralino has relatively large freeze-out density for most of the parameter space.

\subsection{Stau NLSP}

If produced, a stau NLSP would be seen as a massive stable charged particle at colliders. It would leave a clean track in the inner detector and then reach the muon detector, hence it would look like a slow/heavy muon. Because of its electromagnetic charge, the stau can be slowed down by making it go through a bulky medium. Thus it can be trapped and stored until it decays\cite{FS,HNR}.
In this way one can hope to measure its lifetime. 

Within the CMSSM, stau NLSP has the largest allowed region of parameter space, hence thought as the natural candidate. Stau NLSP would yield catalytic effect on BBN. This has attract much attention and many papers are devoted in the study of this topic, in particular regarding the lithium problem solution\cite{stauCat}.

\subsection{Stop NLSP}

A long lived stop would hadronize once it is produced. By taking analogy with heavy quark hadrons one can deduce the lightest hadron states and their lifetimes. The light stop sbaryons are $\Lambda_{\widetilde T}^+ \equiv {\tilde t_1} ud$ (which is the lightest), $\Sigma_{\widetilde T}^{++,+,0} \equiv \tilde{t}_1 (uu, ud, dd)$ (which decays through strong interaction), and $\Xi_{\widetilde T}^{+,0} \equiv \tilde{t}_1 s (u,d)$ (which decays semileptonically with lifetime $\tau \lesssim 10^{-2}$~s). The light stop mesinos are ${\widetilde T}^0 \equiv \tilde{t}_1 {\bar u}$ (which is the lightest), ${\widetilde T}^+ \equiv \tilde{t}_1 {\bar d}$ (with lifetime $\tau \simeq 1.2$~s), and ${\widetilde T}_s \equiv \tilde{t}_1 {\bar s}$ (with lifetime $\tau \simeq 2 \times 10^{-6}$~s). 
The antistop would hadronize into the corresponding antisbaryons and antimesinos.
In the early universe, being the lighter one, the neutral $\widetilde{T}^0$ is more abundance than the charged $\Lambda_{\widetilde T}^\pm$. This reduces the catalytic effect on BBN. Moreover, due to its strong interaction, the freeze out density of stop is generally small. Therefore this scenario can generally satisfy the BBN constraint\cite{stopNLSP}.

On the other hand, it was shown\cite{KS} that stop NLSP scenario is fit to solve the lithium problem through hadronic decay. Note that further annihilation of stop occurs after the hadronization, with annihilation rate $\Gamma_{\rm ann} = \langle \sigma v \rangle n_{\tilde t}$ where $\sigma \sim R_{\rm had}^2$ and $v \simeq \sqrt{3 T / m_{\tilde t}}$. 
The final stop abundance before its decay can be written as
\begin{eqnarray}
m_{\tilde{t}} Y_{\tilde{t}} 
    &=& 0.87 \times 10^{-14} \, {\rm GeV} \left(
    \frac{f_{\sigma}}{0.1} \right)^{-2} \left( \frac{g_{*}}{17.25}
    \right)^{-1/2} \nonumber \\ 
&& \times \left( \frac{T_{\rm QCD}}{150 \, {\rm MeV}}
    \right)^{-3/2} \left( \frac{m_{\tilde{t}}}{10^{2} \, {\rm
    GeV}} \right)^{3/2}
\end{eqnarray}
It was found that, with $f_\sigma = 0.1$, the lithium problem solution prefers $m_{\tilde{t}} = 400 - 600$~GeV and $m_{3/2} = 2-10$~GeV.

\subsection{Sneutrino NLSP}

Since sneutrino does not interact strongly nor electromagnetically the effect of sneutrino decays on BBN can be guessed to be small, but nonzero. The BBN effect comes through energy transfer (elastic/inelastic) from energetic neutrino (produced by the decay $\tilde{\nu} \to \widetilde{G} + \nu$) to the background particles, and from 3/4-body decay modes\cite{KKKM} ($\tilde{\nu} \to \widetilde{G} + \nu + (\gamma, Z)$, $\tilde{\nu} \to \widetilde{G} + \ell + W$) and 4-body ($\tilde{\nu} \to \widetilde{G} + \nu + f + \bar{f}$,
$\tilde{\nu} \to \widetilde{G} + \ell + f + \bar{f}^\prime$). Although the 3 and 4 body decay branching ratios are small, they can still be important since they produced particles that can directly involve in the nucleosynthesis processes. 
For $M_{\tilde{\nu}} \sim O(100$~GeV), the BBN constraint can be satisfied if the sneutrino density before the decay is   
\begin{eqnarray}
Y_{\tilde{\nu}} M_{\tilde{\nu}} &\lesssim & \mathcal{O}(10^{-11}) \ {\rm GeV} \hspace{1cm} {\rm for} \quad  B_h =
10^{-3} \\
Y_{\tilde{\nu}} M_{\tilde{\nu}} &\lesssim & \mathcal{O}(10^{-8}) \ {\rm GeV} \hspace{1cm} {\rm for} \quad  B_h = 10^{-6}
\end{eqnarray}
where $B_h$ is the hadronic branching ratio. The sneutrino NLSP and gravitino LSP scenario was explored within NUHM models, and it was found that there are large regions of parameter space still allowed\cite{GDM-NUHM}.

At colliders, similar to the neutralino (N)LSP case, sneutrino NLSP would yield a missing energy signature. We can study this scenario through cascade decays of heavier supersymmetric particles. In general, the signatures are different from those in neutralino case. Signatures that are thought to be the best for neutralino LSP might not be suitable for this sneutrino NLSP case. A preliminary study of collider phenomenology with sneutrino NLSP has been done in Ref.~\refcite{CoKra}. However, a more detail study might reveal more information.

\section{Concluding Remarks}

Gravitino is a feasible and interesting candidate for dark matter.
There are many possible phenomenology with gravitino dark matter depending
on the choice for the NLSP, which we have summarized in this article.  
Future and upcoming collider experiments, such as the LHC, might be able to unveil some hints on the identity of dark matter. Progresses in direct and indirect detection experiments are also looked promising. The next few years would be an interesting time to find out more about dark matter, and whether gravitino can still stand up as a candidate for dark matter.

\section*{Acknowledgments}
My participation in Dark2009 conference was made possible by the support of British Royal Society. I thank my collaborators on the subject of this paper: John Ellis, Lorenzo Diaz-Cruz, Terrance Figy, Kazunori Kohri, Keith Olive, Krzysztof Rolbiecki, and Vassilis Spanos.


\begin{thebibliography}{9}

\bibitem{EHNOS}
  J.~R.~Ellis, J.~S.~Hagelin, D.~V.~Nanopoulos, K.~A.~Olive and M.~Srednicki,
  Nucl.\ Phys.\  B {\bf 238}, 453 (1984).

\bibitem{snuL}
  T.~Falk, K.~A.~Olive and M.~Srednicki,
  Phys.\ Lett.\  B {\bf 339}, 248 (1994).

\bibitem{GDM-CMSSM}
  J.~R.~Ellis, K.~A.~Olive, Y.~Santoso and V.~C.~Spanos,
  Phys.\ Lett.\  B {\bf 588}, 7 (2004).

\bibitem{otherGDM}
  J.~L.~Feng, A.~Rajaraman and F.~Takayama,
  Phys.\ Rev.\ Lett.\  {\bf 91}, 011302 (2003);
  Phys.\ Rev.\  D {\bf 68}, 063504 (2003);
  J.~L.~Feng, S.~f.~Su and F.~Takayama,
  Phys.\ Rev.\  D {\bf 70}, 063514 (2004);
  Phys.\ Rev.\  D {\bf 70}, 075019 (2004);
  L.~Roszkowski, R.~Ruiz de Austri and K.~Y.~Choi,
  JHEP {\bf 0508}, 080 (2005).



\bibitem{SUGRA}
  A.~H.~Chamseddine, R.~L.~Arnowitt and P.~Nath,
  Phys.\ Rev.\ Lett.\  {\bf 49}, 970 (1982);
  R.~Barbieri, S.~Ferrara and C.~A.~Savoy,
  Phys.\ Lett.\  B {\bf 119}, 343 (1982);
  L.~J.~Hall, J.~D.~Lykken and S.~Weinberg,
  Phys.\ Rev.\  D {\bf 27}, 2359 (1983).

\bibitem{GMSB}
  G.~F.~Giudice and R.~Rattazzi,
  Phys.\ Rept.\  {\bf 322}, 419 (1999).

\bibitem{moroi}
  T.~Moroi,
  arXiv:hep-ph/9503210.


\bibitem{NUHM}
  D.~Matalliotakis and H.~P.~Nilles,
  Nucl.\ Phys.\  B {\bf 435}, 115 (1995);
  M.~Olechowski and S.~Pokorski,
  Phys.\ Lett.\  B {\bf 344}, 201 (1995);
  V.~Berezinsky, A.~Bottino, J.~R.~Ellis, N.~Fornengo, G.~Mignola and S.~Scopel,
  Astropart.\ Phys.\  {\bf 5}, 1 (1996);
  M.~Drees, M.~M.~Nojiri, D.~P.~Roy and Y.~Yamada,
  Phys.\ Rev.\  D {\bf 56}, 276 (1997)
  [Erratum-ibid.\  D {\bf 64}, 039901 (2001)];
  P.~Nath and R.~L.~Arnowitt,
  Phys.\ Rev.\  D {\bf 56}, 2820 (1997);
  A.~Bottino, F.~Donato, N.~Fornengo and S.~Scopel,
  Phys.\ Rev.\  D {\bf 63}, 125003 (2001);
  S.~Profumo,
  Phys.\ Rev.\  D {\bf 68}, 015006 (2003);
  D.~G.~Cerdeno and C.~Munoz,
  JHEP {\bf 0410}, 015 (2004);
  H.~Baer, A.~Mustafayev, S.~Profumo, A.~Belyaev and X.~Tata,
  JHEP {\bf 0507}, 065 (2005);
  J.~R.~Ellis, K.~A.~Olive and P.~Sandick,
  Phys.\ Rev.\  D {\bf 78}, 075012 (2008);
  L.~Roszkowski, R.~R.~de Austri, R.~Trotta, Y.~L.~Tsai and T.~A.~Varley,
  arXiv:0903.1279 [hep-ph].


\bibitem{ourNUHM}
  J.~R.~Ellis, K.~A.~Olive and Y.~Santoso,
  Phys.\ Lett.\  B {\bf 539}, 107 (2002);
  J.~R.~Ellis, T.~Falk, K.~A.~Olive and Y.~Santoso,
  Nucl.\ Phys.\  B {\bf 652}, 259 (2003).


\bibitem{stopNLSP}
  J.~L.~Diaz-Cruz, J.~R.~Ellis, K.~A.~Olive and Y.~Santoso,
  JHEP {\bf 0705}, 003 (2007).

\bibitem{GDM-NUHM}
  J.~R.~Ellis, K.~A.~Olive and Y.~Santoso,
  JHEP {\bf 0810}, 005 (2008). 


\bibitem{ELN}
  J.~R.~Ellis, A.~D.~Linde and D.~V.~Nanopoulos,
  Phys.\ Lett.\  B {\bf 118}, 59 (1982).

\bibitem{KL}
  M.~Y.~Khlopov and A.~D.~Linde,
  Phys.\ Lett.\  B {\bf 138}, 265 (1984).

\bibitem{graInf}
  M.~Kawasaki, F.~Takahashi and T.~T.~Yanagida,
  Phys.\ Lett.\  B {\bf 638}, 8 (2006).



\bibitem{BBB}
  M.~Bolz, A.~Brandenburg and W.~Buchmuller,
  Nucl.\ Phys.\  B {\bf 606}, 518 (2001)
  [Erratum-ibid.\  B {\bf 790}, 336 (2008)].

\bibitem{WMAP5}
  E.~Komatsu {\it et al.}  [WMAP Collaboration],
  Astrophys.\ J.\ Suppl.\  {\bf 180}, 330 (2009).

\bibitem{TR}
  J.~Pradler and F.~D.~Steffen,
  Phys.\ Lett.\  B {\bf 648}, 224 (2007);
  K.~Y.~Choi, L.~Roszkowski and R.~Ruiz de Austri,
  JHEP {\bf 0804}, 016 (2008).


\bibitem{pospelov}
  M.~Pospelov,
  Phys.\ Rev.\ Lett.\  {\bf 98}, 231301 (2007);
  C.~Bird, K.~Koopmans and M.~Pospelov,
  Phys.\ Rev.\  D {\bf 78}, 083010 (2008).

\bibitem{otherCat}
  K.~Kohri and F.~Takayama,
  Phys.\ Rev.\  D {\bf 76}, 063507 (2007);
  M.~Kaplinghat and A.~Rajaraman,
  Phys.\ Rev.\  D {\bf 74}, 103004 (2006).

\bibitem{beryCat}
  M.~Pospelov, J.~Pradler and F.~D.~Steffen,
  JCAP {\bf 0811}, 020 (2008).

\bibitem{jedamzik}
  K.~Jedamzik,
  Phys.\ Rev.\  D {\bf 70}, 063524 (2004);
  D.~Cumberbatch, K.~Ichikawa, M.~Kawasaki, K.~Kohri, J.~Silk and G.~D.~Starkman,
  Phys.\ Rev.\  D {\bf 76}, 123005 (2007).




\bibitem{lamon+durrer}
  R.~Lamon and R.~Durrer,
  Phys.\ Rev.\  D {\bf 73}, 023507 (2006).

\bibitem{COBE-FIRAS}
  D.~J.~Fixsen, E.~S.~Cheng, J.~M.~Gales, J.~C.~Mather, R.~A.~Shafer and E.~L.~Wright,
  Astrophys.\ J.\  {\bf 473}, 576 (1996).

\bibitem{warmgravitino}
  M.~Kaplinghat,
  Phys.\ Rev.\  D {\bf 72}, 063510 (2005);
  J.~A.~R.~Cembranos, J.~L.~Feng, A.~Rajaraman and F.~Takayama,
  Phys.\ Rev.\ Lett.\  {\bf 95}, 181301 (2005);
  F.~D.~Steffen,
  JCAP {\bf 0609}, 001 (2006).



\bibitem{fairbairn}
  M.~Fairbairn {\it et al.},
  Phys.\ Rept.\  {\bf 438}, 1 (2007).

\bibitem{CDF-CHAMPS}
  T.~Aaltonen {\it et al.}  [CDF Collaboration],
  arXiv:0902.1266 [hep-ex].

\bibitem{D0-CHAMPS}
  V.~M.~Abazov {\it et al.}  [D0 Collaboration],
  arXiv:0809.4472 [hep-ex].

\bibitem{CDF-chi}
  T.~Aaltonen {\it et al.}  [CDF Collaboration],
  Phys.\ Rev.\  D {\bf 78}, 032015 (2008).

\bibitem{CoKra}
  L.~Covi and S.~Kraml,
  JHEP {\bf 0708}, 015 (2007).

\bibitem{charginoNLSP}
  G.~D.~Kribs, A.~Martin and T.~S.~Roy,
  JHEP {\bf 0901}, 023 (2009).


\bibitem{FS}
  J.~L.~Feng and B.~T.~Smith,
  Phys.\ Rev.\  D {\bf 71}, 015004 (2005)
  [Erratum-ibid.\  D {\bf 71}, 019904 (2005)].


\bibitem{HNR}      
        K.~Hamaguchi, M.~M.~Nojiri and A.~de Roeck,
        JHEP {\bf 0703}, 046 (2007).


\bibitem{stauCat}
  D.~G.~Cerdeno, K.~Y.~Choi, K.~Jedamzik, L.~Roszkowski and R.~Ruiz de Austri,
  JCAP {\bf 0606}, 005 (2006);
  R.~H.~Cyburt, J.~R.~Ellis, B.~D.~Fields, K.~A.~Olive and V.~C.~Spanos,
  JCAP {\bf 0611}, 014 (2006);
  J.~Kersten and K.~Schmidt-Hoberg,
  JCAP {\bf 0801}, 011 (2008);
  J.~Pradler and F.~D.~Steffen,
  Phys.\ Lett.\  B {\bf 666}, 181 (2008);
  M.~Kawasaki, K.~Kohri, T.~Moroi and A.~Yotsuyanagi,
  Phys.\ Rev.\  D {\bf 78}, 065011 (2008);
  G.~Panotopoulos,
  Phys.\ Lett.\  B {\bf 671} 327 (2009);
  S.~Bailly, K.~Jedamzik and G.~Moultaka,
  arXiv:0812.0788 [hep-ph].




\bibitem{KS}
  K.~Kohri and Y.~Santoso,
  arXiv:0811.1119 [hep-ph].


\bibitem{KKKM}
  T.~Kanzaki, M.~Kawasaki, K.~Kohri and T.~Moroi,
  Phys.\ Rev.\  D {\bf 76}, 105017 (2007).




\end{thebibliography}
\end{document}